# Robust Room-Temperature Quantum Spin Hall Effect in Methyl-functionalized InBi honeycomb film

Sheng-shi Li,[a,b] Wei-xiao Ji,[a] Chang-wen Zhang*,[a] Shu-jun Hu,[b] and Ping Li,[a] Pei-ji Wang [a]

[a] School of Physics and Technology, University of Jinan, Jinan, Shandong, 250022, People's Republic of China

[b] School of Physics, State Key laboratory of Crystal Materials, Shandong University, Jinan, Shandong, 250100, People's Republic of China

**Abstract**

Two-dimensional (2D) group-III-V honeycomb films have attracted significant interest for their potential application in fields of quantum computing and nanoeletronics. Searching for 2D III-V films with high structural stability and large-gap are crucial for the realizations of dissipationless transport edge states using quantum spin Hall (QSH) effect. Based on first-principles calculations, we predict that the methyl-functionalized InBi monolayer (InBiCH$_3$) has no dynamic instability, and host a QSH state with a band gap as large as 0.29 eV, exhibiting an interesting electronic behavior viable for room-temperature applications. The topological characteristic is confirmed by $s$-$p_{xy}$ bands inversion, topological invariant $Z_2$ number, and the time-reversal symmetry protected helical edge states. Noticeably, the QSH states are tunable and robust against the mechanical strain, electric field and different levels of methyl coverages. We also find that InBiCH$_3$ supported on $h$-BN substrate maintains a nontrivial QSH state, which harbors the edge states lying within the band gap of substrate. These findings demonstrate that the methyl-functionalized III-V films may be a good QSH platform for device design and fabrication in spintronics.

*Correspondence and requests for materials should be addressed to: zhchwsd@163.com

# I. Introduction

Two-dimensional (2D) topological insulators (TIs), known as quantum spin Hall (QSH) insulators, have attracted significant researches interest in condensed matter physics and materials science [1,2,3]. The unique characteristic of 2D TI is generating the chiral 1D helical edge states inside the bulk-gap, where the states moving in opposite direction have opposite spin. Therefore, the backscattering is prohibited as long as the scattering potential does not break time-reversal symmetry (TRS) and such helical edge states provide a new mechanism to realize non-dissipative electronic transportation, which promises potential application in low-power and multi-functional device [1,2]. This opens a new avenue in the quest for searching and designing TIs in 2D systems. From this point of view, a 2D TI with a large gap, chemically stable under ambient condition exposure to air, and easy to prepare experimentally, is highly desired. [4,5] The prototypical concept of QSH insulator is first proposed by Kane and Mele in graphene [6,7], in which the spin-orbital coupling (SOC) opens gaps at the Dirac point. However, due to the rather weak SOC, the associated gap is extremely small ($\sim 10^{-3}$ meV) [8], which makes the QSH effect can only be observed at an unrealistically low temperature. Fortunately, this unobservable bulk band gap has predicted to be enhanced by constructing van der Waals heterostructures with graphene and other 2D materials which possess strong SOC effect, such as $Bi_2Se_3$[9], $WS2$[10] and chalcogenides BiTeX (X=Cl, Br and I) [11]. Quantized conductance through QSH edge states have been experimentally demonstrated in HgTe/CdTe [12,13] and InAs/GaSb [14,15] quantum wells, demonstrating an interesting in further experimental studies and possible applications.

Currently, searching for new QSH insulators with controllable quantum phase transitions and tunable electronic and spin properties is crucial for promoting TI technology. After the discovery of QSH phase in graphene, silicene, germancene and stanene are authenticated as QSH insulator sequentially[16], in which the band gap in stanene reaches 0.1eV due to its stronger SOC[16,17]. Then the research focus extends to Group-V bilayer films, and the results prove that Bi (111) [18] bilayer is a QSH insulator intrinsically. Recently, the investigation emphasis is transferred to the 2D inversion-asymmetric materials which contain two different elements, such as Group III-V binary compounds[19-21]. According to prediction, suitable bulk band gap and isolated Dirac cone in these 2D III-V compounds make them suitable for room-temperature applications. A computational study[22] has suggested the possible synthesis and potential applications for 2D III-V compounds. Moreover, due to the inversion-asymmetry in III-V films, they may host nontivial topological phenomena, such as crystalline-surface-dependent topological electronic states[23, 24], natural topological *p-n* junctions[25] and topological magnetoelectric effects[26]. Nevertheless, the surface activity is a major problem for these materials, whose topology may be destroyed by substrate. Besides, some 2D transition-metal dichalcogenides[27] and halide[28] are predicted to be a new class of QSH insulator with a large band gaps.

Recent studies have highlighted that the orbital filtering effect (OFE) plays an important role on tuning the bulk band gap, which has received intense attentions in designing of QSH insulators [16 17 29 30 31 32]. The 2D films are advantageous in this aspect since their bonding properties are easy to be modified in the processes of synthesis to enhance SOC by surface adsorption. For instance, the band gap of stanene will be enhanced to 0.3eV with the effect of halogenation[17], and the SOC is confined in the $p_{xy}$ orbitals rather than $p_z$ in pristine one. The 2D BiX/SbX (X = H, F, Cl and Br) films are also predicted as TIs with extraordinarily large band gap from 0.32 eV to a record value of 1.08 eV [29]. Also, the Bi and Sb films have been demonstrated to contain a large band gap using Si substrate as a tool to achieve OFE [30 31]. Recently, OFE is also applicable in materials with inversion-asymmetric sublattices. The Group IV non-centrosymmetric honeycomb lattices functionalized with halogen atoms are theoretically confirmed to be topological insulators with appropriate condition[33]. Meanwhile, for the functionalized III-Bi films, they can preferably realize nontrivial topological states, producing remarkable large Rashba spin splitting effect [32 21]. By the way, through the combination of Si (111) substrate and hydrogenation, it can still implement OFE in III-V compounds films[34]. These QSH insulators are essential for realizing many exotic phenomena and for fabricating new quantum devices that operates at room temperature. Unfortunately, though their band gap has been enhanced greatly, further experimental works [35] find that the fluorination and hydrogenation exhibit quick kinetics, with rapid increasing of defects and lattice disorder even under short plasma exposures, which will disturb their potential applications completely. Therefore, the realization of these QSH insulators with high quality is rather difficult [36,37,38].

Recently, the small molecule functionalization has been the focus to enhance the geometric stability and nontrivial band gap of new 2D films. For instance, the ethynyl ($C_2H$) has been reported to be an efficient way to stabilize stanene by decoration on its surface, and its band gap can be enhanced to 0.3 eV. [39] Methyl ($CH_3$), another organic molecule, has also been suggested as a promising tool to stabilize 2D systems, such as methyl-functionalized germanane ($GeCH_3$) [40] and $BiCH_3$ [41], to realize large gap QSH insulators. Experimentally, $GeCH_3$ film has been synthesized in recent work [42]. In contrast to hydrogenated germanane, the $GeCH_3$ film has considerably enhanced thermal stability at high-temperature. [42] This raises an interesting question: can the methyl be applied to stabilize group III-V films, and whether their band gap can be enhanced significantly in QSH phase?

In this work, we address the aforementioned question by demonstrating the effect of methyl-functionalization as an effective way to stabilize InBi monolayer ($InBiCH_3$) as a representative for group III-V films. Indeed, the methyl removes the imaginary frequency modes from the phonon spectrum of InBi, indicating no dynamic instability. In this case, the band gap of $InBiCH_3$ can be enhanced to 0.29 eV, which is larger than that of pristine InBi monolayer. The physical origin of QSH effect is confirmed by identifying the $s$-$p_{xy}$ bands inversion, topological invariant $Z_2 = 1$, and helical edge states in bulk band gap. Noticeably, the QSH effects are robust and tunable against the mechanical strain, electric field and different level of methyl coverages. In addition, the $InBiCH_3$ on $h$-BN sheet is observed to support a nontrivial large gap QSH, which harbors the edge states lying within the band gap of $h$-BN sheet. Therefore, our work reveals a unique advantage of atomically thin III-V materials in realizing 2D topological phase.

**II. Calculation Details**

First-principles calculations based on density functional theory (DFT) were carried out using the plane wave basis Vienna ab initio simulation package [43,44]. The electron-ion potential was described by the projector-augmented wave (PAW) method [45] and the electron exchange-correlation functional was approximated by generalized gradient approximation (GGA) in Perdew-Burke-Ernzerhof (PBE) form [46]. The energy cutoff of the plane waves was set to 500 eV with the energy precision of $10^{-5}$ eV. The vacuum space was applied at least 20 Å to eliminate the interactions between neighboring slabs in *z*-direction. We employed a k-point set generated by the $11 \times 11 \times 1$ Monkhorst-Pack mesh [47] for geometry optimizations and $17 \times 17 \times 1$ for self-consistent calculations. The atomic coordinates were fully optimized until the force on each atom was less than 0.01 eV/Å. The SOC is included in self-consistent electronic structure calculations. The phonon spectra were calculated using a supercell approach within the PHONON code [48].

**III. Results and Discussion**

Before discussing the electronic properties of the methyl-functionalized InBi, we examine the crystal structure, dynamic stability, and electronic properties of the pristine InBi monolayer. Figure 1(a) presents the geometric structure of InBi, composed of a triangular lattice with In and Bi atoms located in two different sublattices. Our calculations indicate that, in analogy to buckled silicene [49], InBi exhibits a hexagonal structure with a buckling parameter $\Delta = 0.873$ Å, lattice constant $a = 4.78$ Å, and bond length $d = 2.89$ Å, which consist with those reported in previous literature[20,34], but are smaller than the cases in two films of InBi[50] except $\Delta$. The resultant band structure indicates that it is a QSH insulator, with a band gap of 0.16 eV. Such an upper limit, however, can be significantly broken through, as we will show in the following part. In addition, the phonon spectrum calculations in Figure 1(c) indicate that it has obviously imaginary frequency modes, exhibiting a dynamically unstable structure.

To stabilize InBi monolayer, we saturate the uncoordinated In and Bi atoms with methyl alternating on both sides of InBi sheet, as shown in Figure 1(b). In comparison with InBi monolayer, the lattice constant of InBiCH$_3$ is stretched to $a$ = 4.89 Å upon methyl-functionalization, along with the Ga−Bi bond length increasing by 0.08 Å. Noticing that this lattice constant is larger than hydrogenated case[34], which will enhance the localization of atom orbitals, the methyl-functionalized InBi film may process a more favorable character than the hydrogenated one. Meanwhile, the In−CH$_3$ and Bi−CH$_3$ bonds are 2.23 and 2.30 Å, respectively. Importantly, the stability of 2D InBiCH$_3$ is confirmed by phonon spectrum that clearly removes the imaginary frequencies from pristine InBi monolayer, as displayed in Figure 1(d). Also, we calculate the formation energy of InBiCH$_3$ defined as:

$$E_f = E(\text{InBiCH}_3) - [E(\text{InBi}) + E(\text{CH}_3)] \qquad (1)$$

where $E(\text{InBiCH}_3)$ and $E(\text{InBi})$ are total energies of InBiCH$_3$ and InBi, respectively, while $E(\text{CH}_3)$ is chemical potential of methyl. It is found to be $E_f$ = -3.23 eV, greatly lager than the cases of GaBiCl$_2$ [51] and GeCH$_3$ (-1.75 eV). These indicate that the methyl strongly binds to InBi monolayer by a chemical bonds, showing a higher thermodynamic stability relative to their elemental reservoirs. Considering that the GeCH$_3$ has been successfully synthesized, [42] InBiCH$_3$ is also expected to be feasible experimentally.

Figure 2 displays the calculated band structures in InBiCH$_3$ monolayer. In the absence of SOC, it is a semiconductor with a direct band gap of 0.21 eV at the Γ point, as shown in Figures 2(a-b). When SOC is switched on, there still has a direct band gap ($E_\Gamma$) of 0.31eV at the Γ point, which is twice larger than that of pristine one. However, the valence band minimum (VBM) moves slightly away from the Γ point, leading to an indirect band gap ($E_g$) of 0.29 eV [Figures 2(c-d)], which is significantly larger than the InBi film with hydrogenation(0.19eV) [34], verifying the assumption proposed above. The SOC-induced band structure deformation near the Fermi level is a strong indication of the existence of topologically nontrivial phase. To further reveal the effect of chemical decoration of methyl, we project the bands onto different atomic orbitals. The energetically degenerated VBM without SOC are mainly derived from $p_{xy}$ orbitals, whereas the conduction band minimum (CBM) is contributed by $s$ orbital [Figure 2(a)]. It is known that the $s$ orbital locates typically above $p$ orbitals in conventional III-V compounds. Consequently, the InBiCH$_3$ exhibits a normal band order. However, the effect of SOC makes $s$ and $p_{xy}$ components at the Γ point exchanged, resulting in an inverted band order, as shown in Figure 2(c). Here, there is a nontrivial bulk band gap of 0.29 eV at the Fermi level, considerably exceeding the presumed upper limits settled by the system without decoration. Here, we highlight that the QSH states of InBiCH$_3$ are markedly different from ethynyl- or methyl-functionalized stanene [39] and germanene [40], which are trivial TIs at the equilibrium state. In addition, due to the inversion-symmetry breaking in InBiCH$_3$, we also find that the resulting band structure is different from stanene film, [16] and shows intriguing Rashba-type dispersions, as shown in Figures 2(c) and (d). This spin splitting is also similar to what occurs for the hydrogenated InBi monolayer. [20]

The most important performance for TIs is the existence of helical edge states with spin polarization protected by TRS, which can be calculated by the Wannier 90 package [48]. Based on maximally localized Wannier functions (MLWFs), the edge Green's function [52] of the semi-infinite lattice is constructed using the recursive method, and the local density of state (LDOS) of the edges is presented in Figure 3(a). We can see that all the edge bands connect completely the conduction and valence bands and span the 2D bulk band gap, yielding a 1D gapless edge states. Further, by identifying the spin-up (↑) and spin-down (↓) contributions in the edge spectral function [Figure 3(b)], the counter-propagating edge states can exhibit opposite spin-polarization, in accordance with the spin-momentum locking of 1D helical electrons. Furthermore, the Dirac point located at the band gap are calculated to have a high velocity of ~ $2.0 \times 10^5$ m/s, comparable to that of $5.3 \times 10^5$ m/s in HgTe/CdTe quantum well [12] [13]. All these results consistently demonstrate that the InBiCH$_3$ is an ideal 2D TI.

The topological states can be further confirmed by calculating topological invariant $Z_2$. Due to the inversion-symmetry breaking in InBiCH$_3$, the method proposed by Fu and Kane [53] cannot be used, and thus, an alternative one independent of the presence of inversion-symmetry is needed. Here, we introduce the evolution of Wannier Center of Charges (WCCs) [54] to calculate the $Z_2$ invariant, which can be expressed as:

$$Z_2 = P_\theta(T/2) - P_\theta(0) \quad (2)$$

which indicates the change of time-reversal polarization ($P_\theta$) between the 0 and $T/2$. Then the WFs related with lattice vector R can be written as :

$$|R,n\rangle = \frac{1}{2\pi}\int_{-\pi}^{\pi} dk e^{-ik(R-x)}|u_{nk}\rangle \quad (3)$$

Here, a WCC $\bar{x}_n$ can be defined as the mean value of $\langle 0n|\hat{X}|0n\rangle$, where the $\hat{X}$ is the position operator and $|0n\rangle$ is the state corresponding to a WF in the cell with $R = 0$. Then we can obtain

$$\bar{x}_n = \frac{i}{2\pi}\int_{-\pi}^{\pi} dk \langle u_{nk}|\partial_k|u_{nk}\rangle \quad (4)$$

Assuming that $\sum_\alpha \bar{x}_\alpha^S = \frac{1}{2\pi}\int_{BZ} A^S$ with $S = I$ or $II$, where summation in $\alpha$ represents the occupied states and $A$ is the Berry connection. So we have the format of $Z_2$ invariant:

$$Z_2 = \sum_\alpha [\bar{x}_\alpha^I(T/2) - \bar{x}_\alpha^{II}(T/2)] - \sum_\alpha [\bar{x}_\alpha^I(0) - \bar{x}_\alpha^{II}(0)] \quad (5)$$

The $Z_2$ invariant can be obtained by counting the even or odd number of crossings of any arbitrary horizontal reference line.

Figure 3(c) displays the evolution lines of WCCs calculated for InBiCH$_3$. One can see that the WCCs evolution curves cross any arbitrary reference lines odd times, which indicates $Z_2 = 1$, verifying the existence of topologically nontrivial phase in InBiCH$_3$.

Strain engineering is an efficient way of modulating the electronic and topological properties in 2D materials [55][56]. We employ in-plane strains to InBiCH$_3$ by changing the lattices as $\varepsilon = (a-a_0)/a_0$, where $a$ ($a_0$) is lattice constant under the strain (equilibrium) condition. Figure 4(a) presents the variation of band gap ($E_g$, $E_\Gamma$) as a function of the biaxial strain. One can see that the nontrivial QSH phase survives in InBiCH$_3$ over a wide range of strains. Under tensile strain, the $E_\Gamma$ enlarges monotonically, and reaches a maximum of 0.79 eV at 20 %. While in the compressive case, the $E_g$ and $E_\Gamma$ is almost consistent with each other, and the $s$-$p_{xy}$ inversion maintains beyond critical point -7 %. If the compressive strain keeps increasing, the trivial band order occurs, forming a normal insulator (NI). The bands inversion characteristics with respect to the strain are illustrated in the insert of Figure 4(a). Here, we must point out that the crystal deformation occurs clearly with relatively large strains, suggesting a robust of QSH effect against crystal deformation.

To elucidate the origin of band topology, we analyze the orbital evolution of InBiCH$_3$ and the results are schematically presented in Figure 4. Since the methyl hybridizes strongly with $p_z$ orbital of In and Bi atoms overlapping in the same energy range, it effectively removes $p_z$ bands away from the Fermi level, leaving only the $s$ and $p_{x,y}$ orbitals at the Fermi level. As shown in Figures 4(c) and (d), the chemical bonding between In and Bi atoms make the $s$ and $p_{x,y}$ orbital split into the bonding and anti-bonding states, *i.e.*, $|s^{\pm}\rangle$ and $|p^{\pm}_{x,y}\rangle$, which the superscripts + and − represent the parities of corresponding states, respectively. In the absence of SOC, the bands near the Fermi level are mainly contributed by the $|s^-\rangle$ and $|p_{x,y}^+\rangle$, with the $|s^-\rangle$ locating above $|p_{x,y}^+\rangle$, possessing a normal band order in the sequence of $s$-$p$. After taking SOC into account, the $|p_{x,y}^+\rangle$ will further split into $|p_{\pm 3/2}^+\rangle$ and $|p_{\pm 1/2}^+\rangle$, while the $|p_{\pm 3/2}^+\rangle$ is pushed up and $|p_{\pm 1/2}^+\rangle$ is shifted down in energy. The hopping between those atomic orbitals plays a vital role in tuning the splitting strength of $|p_{\pm 3/2}^+\rangle$ and $|p_{\pm 1/2}^+\rangle$. In the case of $\varepsilon < -7\%$, the compressive strain leads to a shorter bond length, which increases the splitting degree of bonding and antibonding states, generating a large energy difference between $|p_{x,y}^+\rangle$ and $|s^-\rangle$. Thus, the $|p_{\pm 3/2}^+\rangle$ cannot inverse with $|s^-\rangle$, indicating a trivial band order like conventional III-V compounds, as shown in Figure 4(c). While for $\varepsilon > -7\%$, a smaller energy separation introduced will be arisen and the SOC effect can easily promote the $|p_{\pm 3/2}^+\rangle$ higher than $|s^-\rangle$ [Figure 4(d)], leading to an extraordinary bands inversion order in the sequence of *p-s-p*, namely inversion of parities, indicating the existence of QSH phase. Moreover, to further confirm the nontrivial topological properties, we calculate the edge states when the strain is 105%. The result is presented in Figure S1(a-b), and a pair of helical edge states can be observed obviously, indicating that it still maintains the QSH states except the change of bulk band gap compared with the equilibrium state. It is worth mentioning that this topological states originate from the s-p inversion mechanism, which is common in such III-V compounds rather than p-d or d-d bands inversion in transition-metal compounds[27, 28]. But it is still different from the case of InBi film with fluoridation[32], in which the $|p_{x,y}^+\rangle$ and $|s^-\rangle$ orbitals have inversed in the progress of chemical bonding and the bulk band gap is determined by the splitting of $|p_{x,y}^+\rangle$ under the SOC effect, the bulk band gap in InBiCH3 film is the combined effect of functionalization and SOC, thus yields a smaller bulk band gap.

In addition to strain engineering, the effects of a perpendicular electric field (E-filed) on band topology are investigated for $InBiCH_3$. We find that both $E_g$ and $E_\Gamma$ possess a near-linear dependency with respect to E-field, as illustrated in Figure 4(b). The increases of the positive E-field leads to a larger band gap, while the negative E-field will make it decrease monotonically. More importantly, different from 1T'-MoS2[27] whose topological state will be destroyed by electric field, the $InBiCH_3$ maintains its nontrivial TI nature all the way, indicating a robustness against E-field in the range of −1−1V/ Å. Meanwhile, as a representative, the edge state of 0.5V/ Å is shown in Figure S1(c-d), which is analogous to the case of strain [Figure S1(a-b)]. The nontrivial bulk band gaps are still very large (0.24−0.35 eV), allowing for viable applications at room temperature.

As discussed above, the methyl functionalization does not alter the band topology of InBi monolayer. Thus, we further examine the robustness of nontrivial TI phase by considering different levels of methyl coverages in $InBiCH_3$. A 2×2 supercell is selected to simulate coverages of 0.25, 0.50 and 0.75 monolayer by removing/adding up to methyl molecules. Followed the previous work [57], the methyl molecules are positioned on opposite sides of InBi sheet, yielding a greater stability. Figure 5 presents the relaxed structures and corresponding band structures for 0.25−0.75 monolayer coverage, respectively. Interestingly, all the methyl molecules are strongly hybridized with $p_z$ orbital of In and Bi atoms, thus the atomic states near the Fermi level is still mainly determined by the $s$ and $p_{xy}$ orbitals. Meanwhile, the $s$-$p$ bands inversion can be observed, thus the band topology of these films is highly robust against chemical bonding effects of the environment, making these films particularly flexible to substrate choices for device applications. Besides, we also calculate the phonon spectrums of these different covered monolayers, as shown in Figure S2. According to the results, these structures still possess little imaginary frequency. So the best insurance method is functionalized the InBi film with methyl completely. However, even if it emerges with unsaturated position, it can still own the QSH state, which is a good guarantee. To further verify its robustness against the coverage, taking 0.25 monolayer as a representative, we straightforward construct a ribbon with zigzag edges. The width of this ribbon is 73.53 Å, which is large enough to ignore the effect between two sides. Meanwhile, to eliminate the interaction induced by periodicity, a sufficient vacuum slab is adopted and the edges are passivated by hydrogen atoms. Figure S3(a) presents the corresponding band structure. Different from the results of semi-infinite lattice, two pairs of helical edge states span the bulk energy gap connecting the valence and conduction bands. Due to the asymmetry for edges in zigzag ribbon, a pair of edge bands is determined by the In termination and the other is derived from the Bi termination, also they form two Dirac cones simultaneously at the M point. Noticeably, each pair of edge state crosses the Fermi level with odd number along the direction form M to Γ point which can adequately prove its nontrivial topological properties.

From the view of devices applications, selecting a suitable substrate for $InBiCH_3$ is a very important issue. Considering that one film of InBi with hydrogenation deposited on Si(111) will annihilate its nontrivial topological states[50], we hope to achieve it by constructing a van der Waals heterostructure for its growth. However, the *h*-BN has been reported to be an ideal substrate for 2D materials [58,59], due to its large band gap with a high dielectric constant. Thus, we construct an InBiCH$_3$@2×2*h*-BN heterobilayer (HBL), as shown in Figure 6(a), in which the lattice mismatch is only 0.41% for both layers. After a full relaxation, the distances between adjacent layers are 3.68 Å and 3.05 Å, respectively, with a binding energy of -71 meV, indicating the InBiCH$_3$ interacts weakly with *h*-BN substrate *via* van der Waals interaction [60]. Figure 6(b) presents the band structure of InBiCH$_3$@2×2*h*-BN HBL with SOC. In these weakly coupled system, there is essentially no charge transfer between the adjacent layers, and the HBL remains semiconducting property. By projecting the band structure, we find that the contributions of *h*-BN substrate locate far away from the Fermi level, the states around the Fermi level being dominantly determined by InBiCH$_3$ with an inverted band order. In addition, based on aforesaid method about validation for edge states, we also investigate the band structure of InBiCH$_3$@2×2*h*-BN ribbon. The edge states and odd number of crossing further verify the existence of nontrivial topological property and its robust against substrate, as illustrated in Figure S3(b). If we compare the bands of InBiCH$_3$ with and without *h*-BN substrate, little difference can be observed, which is analogous to the case of tetragonal Bi film on NaCl substrate[61]. As a result, the InBiCH$_3$@2×2*h*-BN HBL is a robust QSH phase with a gap of 0.27 eV.

## IV. Conclusions

In summary, on the basis of first-principles calculations, we have investigated the geometric and electronic properties of InBiCH$_3$. The results indicate that InBiCH$_3$ has no dynamic instability, and is a QSH insulator with a band gap lager than 0.29 eV, suitable for room-temperature application. The topological characteristic can be confirmed by $s$-$p_{xy}$ bands inversion, topological invariant $Z_2 = 1$, and the time-reversal symmetry protected helical edge states. We also find that the band gap of InBiCH$_3$ can be effectively tuned by external strain and electric field. The TI phase is robust against strain (-7 − 20 %) and E-field (-1 − 1V/ Å). Also, the InBiCH$_3$ can preserve nontrivial topology under different levels of methyl coverages. In addition, the InBiCH$_3$ on *h*-BN sheet is observed to support a nontrivial large-gap QSH, which harbors a Dirac cone lying within the band gap. These findings demonstrate that the methyl-functionalized III-V films may be a good QSH platform for device design and fabrication.

——————————

**Acknowledgements**

This work was supported by National Natural Science Foundation of China (Grant Nos. 11274143, 61172028, and 11304121), Natural Science Foundation of Shandong Province (Grant Nos. ZR2013AL004, ZR2013AL002), Technological Development Program in Shandong Province Education Department (Grant No. J14LJ03), Research Fund for the Doctoral Program of University of Jinan (Grant Nos. XBS1433，XBS1402, XBS1452).

**Author contributions**

S. L. and C. Z conceived the study and wrote the manuscript. P. L and P. W. performed the first-principles calculations. W. J calculated the phonon spectrum. S. H., B. Z. and C. C. prepared figures. All authors read and approved the final manuscript.

**Competing financial interests**

The authors declare no competing financial interests.

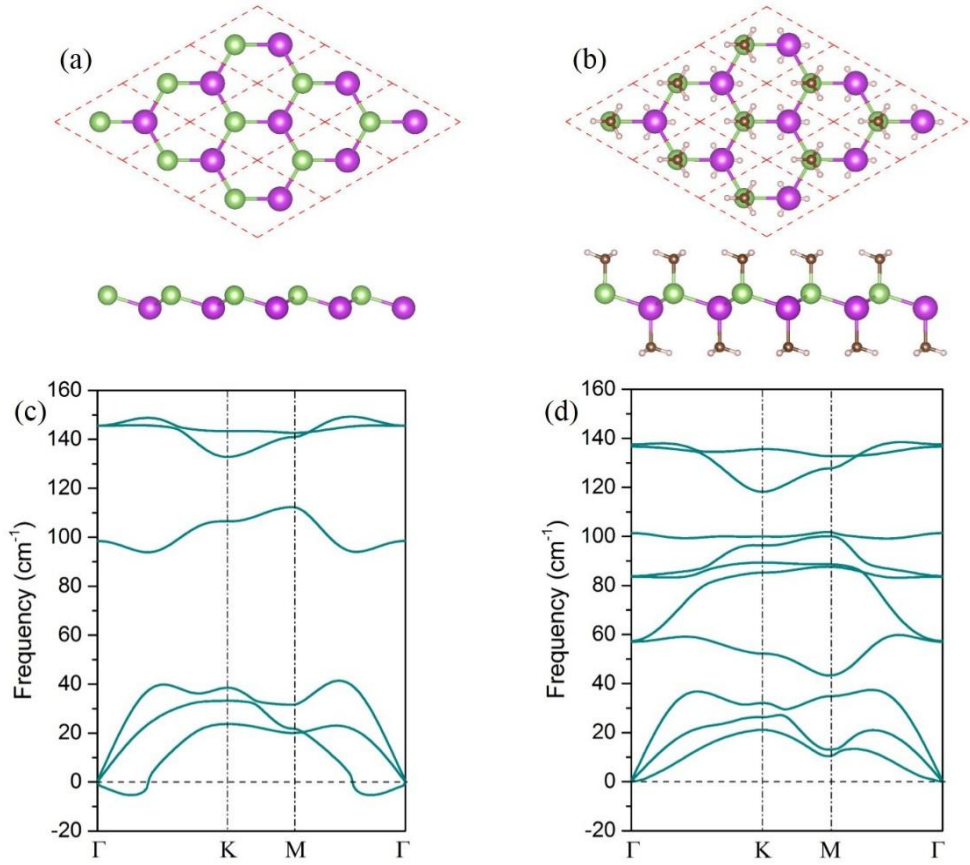

Figure 1. Structural representation (top view and side view) of (a) InBi and (b) InBiCH$_3$. (c,d) Corresponding phonon spectra along the high-symmetric points in the BZ

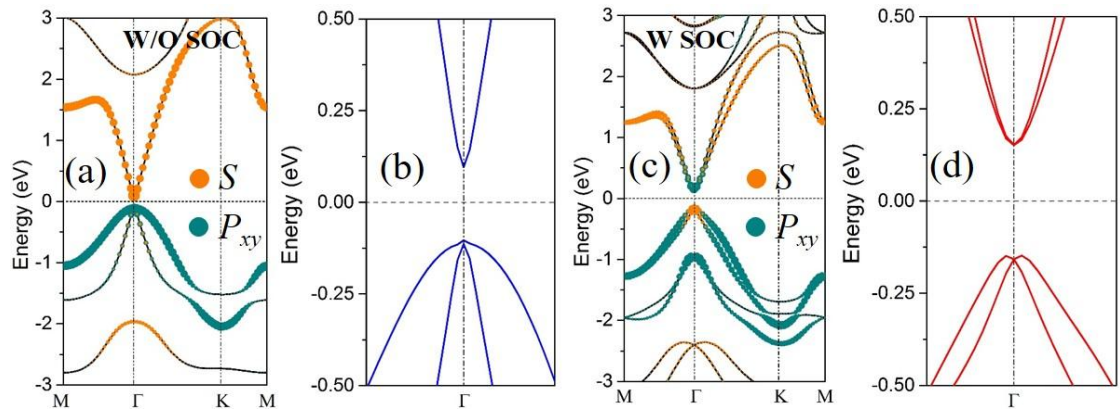

Figure 2. Orbital resolved band structures of InBiCH$_3$ without (a) and with SOC (c), (b) and (d) are the corresponding enlarged views of the bands near the Fermi level at the Γ point. The orange and the light blue dots represent the $s$ and $p_{xy}$ orbitals, respectively.

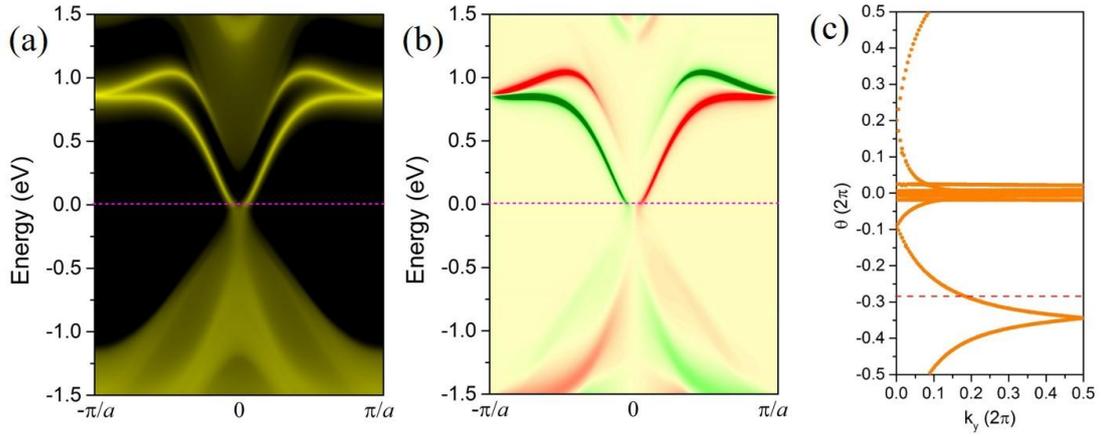

Figure 3. The calculated semi-infinite edge states of total (a) and spin (b) for InBiCH$_3$. The red and green components in (b) denote the spin up and spin down polarization. The Fermi level is set to zero. (c) Evolutions of Wannier centers along $k_y$. The evolution lines cross the arbitrary reference line (red dash line) parallel to $k_y$ odd times, yielding $Z_2 = 1$.

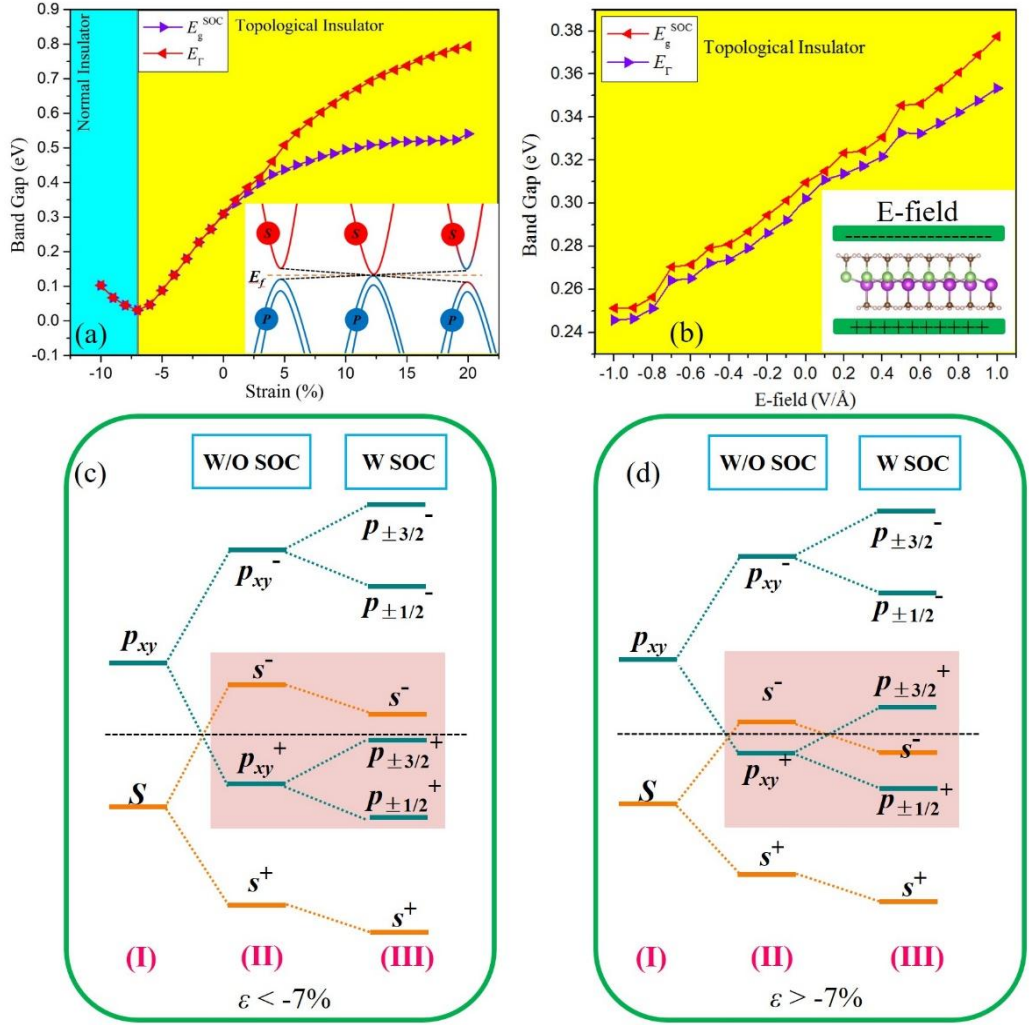

Figure 4. Variation of global gap ($E_g$) and the direct band gap ($E_\Gamma$) as a function of external strain (a) and electric field (b) for InBiCH$_3$. The inserts are the diagram of evolution for atomic orbitals with strain. The evolution of atomic $s$ and $p_{xy}$ orbitals at Γ point is described as chemical bonding and SOC are switched on in sequence for $\varepsilon < -7\%$ (c) and $\varepsilon > -7\%$ (d). The horizontal black dashed lines indicate the Fermi level.

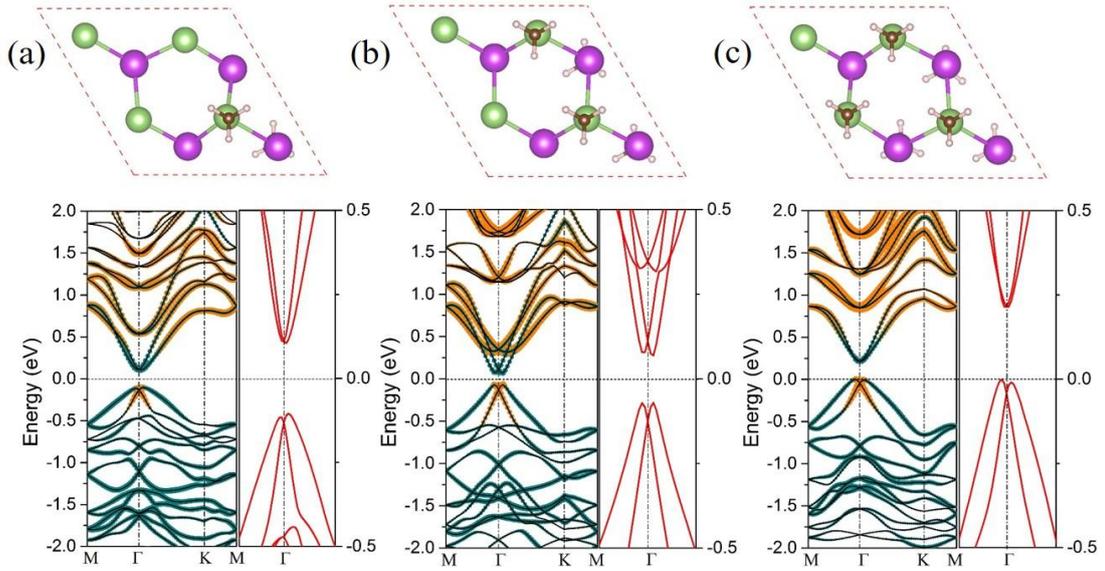

Figure 5. The structure diagram of InBiCH$_3$ monolayer with coverages of (a) 0.25, (b) 0.50 and (c) 0.75. The electronic band structure of (c) 0.25, (b) 0.50 and (c) 0.75 coverages with SOC. The orange and the light blue dots represent the $s$ and $p_{xy}$ orbitals, respectively. The enlarged view of bands near the Fermi level at the Γ point are shown in the right panel.

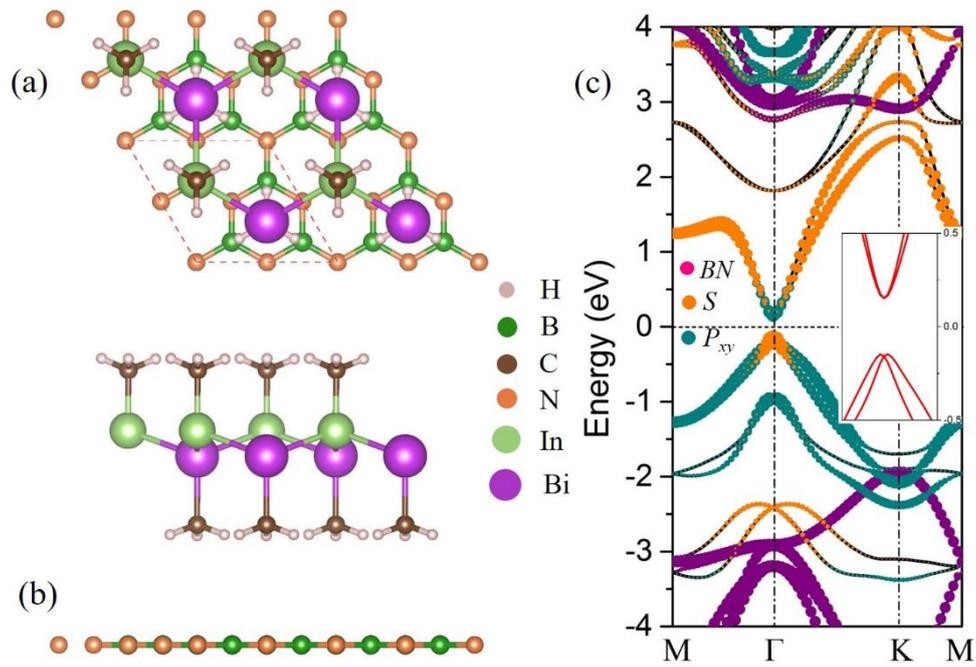

Figure 6. Geometry structure for InBiCH$_3$ deposited on BN substrate from top view (a) and side view (b), as well as orbitals-resolved band structures with SOC (c). The purple dots represent the contribution of BN substrate. The orange and the light blue dots represent the $s$ and $p_{xy}$ orbitals, respectively. The insert is the enlarged drawing of the bands near the Fermi level at the Γ point.